\documentclass{aa}
\usepackage{graphics}
\usepackage{natbib}
\bibpunct{(}{)}{;}{a}{}{,}

\DeclareRobustCommand{\ion}[2]{%
\relax\ifmmode
\ifx\testbx\f@series
{\mathbf{#1\,\mathsc{#2}}}\else
{\mathrm{#1\,\mathsc{#2}}}\fi
\else\textup{#1\,{\mdseries\textsc{#2}}}%
\fi}

\def\apjl{ApJ}%
\def\apjs{ApJS}%
\def\ao{Appl.~Opt.}%
\def\aap{A\&A}%
\title{High-resolution radio continuum survey of M33\\II. Thermal and nonthermal emission}

\author{ F. S. Tabatabaei\inst{1}\thanks{Member of the International Max Planck Research School (IMPRS) for Radio and Infrared Astronomy at the Universities of Bonn and Cologne.}, R. Beck\inst{1}, E. Kr\"ugel\inst{1}, M. Krause\inst{1}, E. M. Berkhuijsen\inst{1}, K. D. Gordon\inst{2}, K.~M. Menten\inst{1}  }
\institute{Max-Planck Institut f\"ur Radioastronomie, Auf dem H\"ugel 69, 53121 Bonn, Germany   
\and Steward Observatory, University of Arizona, 933 North Cherry Avenue, Tucson, AZ 85721
}

\begin{document}

\titlerunning{ High-resolution radio continuum survey of M33}
\authorrunning{Tabatabaei et al.}
\abstract
{ Constraints on the origin and propagation of cosmic rays can be achieved by studying the variation in the spectral index of the synchrotron emission across external galaxies.} 
{To determine the variation in the nonthermal radio spectral index in the nearby spiral galaxy M33 at a linear
resolution of 360\,pc.  }
{ We separate the thermal and nonthermal components of
the radio continuum emission without the assumption of a constant nonthermal
spectral index. Using the Spitzer FIR data at 70 and 160\,$\mu$m and a standard
dust model, we deredden the H$\alpha$ emission.  The extinction corrected
H$\alpha$ emission serves as a template for the thermal free-free radio
emission.  Subtracting from the observed 3.6\,cm and 20\,cm emission (Effelsberg and the VLA) this free-free emission, we obtain the nonthermal maps. 
A constant electron temperature used to obtain the thermal radio intensity seems appropriate for M~33 which, unlike the Milky Way, has a shallow metallicity gradient.
} 
{ For the first time, we derive the distribution of the nonthermal spectral index across a galaxy, M33.  We
detect strong nonthermal emission from the spiral arms and star-forming regions.
Wavelet analysis shows that at 3.6\,cm the nonthermal emission is dominated by
contributions from star-forming regions, while it is smoothly distributed at
20\,cm.  For the whole galaxy, we obtain thermal fractions of 51\% and 18\% at
3.6\,cm and 20\,cm, respectively.  The thermal emission is slightly stronger in
the southern than in the northern half of the galaxy.  We find a clear radial
gradient of mean extinction in the galactic plane.  } 
{ The nonthermal spectral index map indicates that the relativistic electrons suffer energy-loss when
diffusing from their origin in star-forming regions towards interarm regions and
the outer parts of the galaxy. We also conclude that the radio emission is
mostly nonthermal at R\,$>$\,5\,kpc in M33.
%The north-south asymmetry in the distribution of the polarized intensity may be linked to the higher thermal fraction in the southern half of M33. 
%The large exponential scales length of the nonthermal emission hints to a faster 
}
\maketitle

%________________________________________________________________

\section{Introduction}

The problem of separating the two components of the radio continuum emission, free-free (thermal) and synchrotron (nonthermal) emission,  dates back to the beginnings of radioastronomy. The mostly applied techniques are based on either assuming a constant nonthermal spectral index \citep[e.g. ][]{Klein_84} or a correlation between thermal radio and infra-red (IR) emission \citep[][]{Broadbent}.
 
\begin{table*}
\begin{center}
\caption{Images of M33 used in this study }
\begin{tabular}{ l l l l } \hline
Wavelength & Resolution & rms noise &Telescope \\\hline\hline
20\,cm       &  $51\arcsec$ & 70\,$\mu$Jy/beam area &VLA+Effelsberg $^{1}$\\
%20\,cm       &  $540\arcsec$ & Effelsberg $^{2}$\\
3.6\,cm      &  $84\arcsec$ & 230\,$\mu$Jy/beam area &Effelsberg$^{1}$\\
160\,$\mu$m  &  $40\arcsec$ & 10\,$\mu$Jy/arcsec$^2$ &Spitzer$^{2}$\\
70\,$\mu$m  &  $18\arcsec$ & 10\,$\mu$Jy/arcsec$^2$  &Spitzer$^{2}$\\
%24\,$\mu$m  &  $6\arcsec$ & Spitzer$^{3}$\\
6570\AA{}\,(H$\alpha$)   &  $2\arcsec$ (pixel size)& 0.3\,cm$^{-6}$pc  & KPNO$^{4}$ \\
\hline
\noalign {\medskip}
\multicolumn{3}{l}{$^{1}$ \cite{Tabatabaei_1_07}}\\
%\multicolumn{3}{l}{$^{2}$ \cite{Fletcher}}\\
\multicolumn{3}{l}{$^{2}$ \cite{Hinz} and \cite{Tabatabaei_1_07} }\\
\multicolumn{3}{l}{$^{3}$ \cite{Hoopes_et_al_97H}}\\
\end{tabular}
\end{center} 
\label{tab:data}
\end{table*}

Although the assumption of constant nonthermal spectral index may be reasonable for global studies (Sect. 10), it does not lead to a feasible thermal/nonthermal distribution in detailed studies. Under this simplification, it is not possible to investigate the origin and energy loss processes of the electron component of cosmic rays (CRs) within a galaxy.
Discrete synchrotron emitting sources are mainly identified as supernova remnants. The synchrotron emission from supernova remnants can be described as power law ($S_{\nu} \sim \nu^{-\alpha_n}$) with a typical spectral index of $\alpha_n \simeq$~0.6. Propagating from these sources, electrons suffer energy losses that steepens the power law spectrum. In the interstellar medium, the typical spectral index attributed to the emission under the leakage loss is $\simeq$\,0.9, and under synchrotron loss and inverse Compton scattering $\simeq$\,1.1 \citep{Biermann_95}. These electrons further diffuse to the interarm regions and outer parts of spiral galaxies within their life time.  Hence, variations of the nonthermal spectral index  should be distinguishable particularly by comparing arms with interarm regions and outer parts of the galaxies. 

The thermal/nonthermal separation based on the assumption that the radio-IR correlation is due to a correlation between the thermal radio and IR emission is not generally correct, as nonthermal phenomena  e.g. super massive black holes or energetic CRs may stimulate the IR emission. For instance, the possibility of heating the diffuse dust (emitting in IR) by CRs was shown by \cite{Helou_93} and \cite{Becay_90}. Moreover, supernovae, the ultimate sources of most of the nonthermal emission, explode close to the star-forming regions in which their proginator stars formed. Together with the increased magnetic field strength in the spiral arms, this causes a correlation between the thermal and nonthermal sources, making a less direct link between the IR and thermal emission. Finally, the slope of the radio-IR correlation depends on the synchrotron spectral index \citep{Niklas_97}.

Templates for free-free emission could be provided by emission of recombination lines as they originate from within ionized regions, like the free-free emission.  The H$\alpha$ emission, the strongest Balmer line, is observationally most preferred, particularly from nearby galaxies. Both the H$\alpha$ and free-free emission are linearly proportional to the number of ionizing photons produced by massive stars. On the other hand, the H$\alpha$ emission suffers from extinction by dust leading to an underestimate of the free-free emission if no extinction correction is made.  Note that the emission of radio recombination lines is extinction free and hence ideal for tracing the free-free emission. However, these kinds of emission are too weak from the diffuse ionized gas in external galaxies to be detected using the present facilites.  For example, our attempt to map radio recombintion lines emission from diffuse regions within IC~342 and M~33 using the 6.2\,cm receiver of the Effelsberg Telescope was not successful.   

The nearest Scd galaxy, M33\,(NGC\,598), at a distance of 840 kpc \citep[1$\arcsec \simeq$ 4\,pc, ][]{Freedman_etal_91} has been extensively studied at radio and IR wavelengths. With an inclination of $i=56^{\circ}$ \citep{Regan_etal_94},  its spiral structure is well visible. The central position of M33 given by \cite{devaucouleurs_81} is RA(1950)\,=\,$1^{h}31^{m}01.57^{s}$ and  DEC(1950)\,=\,$30^{\circ}24\arcmin15.0\arcsec$. The position angle of the major axis is  PA\,$\simeq 23^{\circ}$ \citep{Deul}. 
So far, extinction studies in M33 have mostly focused on HII regions using either the Balmer (and Paschen) line ratios \citep[e.g. ][]{kwitter, Melnick,Peterson_97} or the ratio of H$\alpha$ to radio emission \citep[e.g. ][]{Israel_80,Devereux_etal_97}. In the line ratio method the emission from the diffuse ionized medium can hardly be detected in most of these recombination lines. The method based on the radio emission only works well when the thermal component of the radio continuum is independently known. For instance, the radio emission from single HII regions \citep[and not HII complexes with a possible nonthermal emission contribution like NGC\,604 and NGC\,595,  ][]{Dodorico_78,Gordons_93} may be considered as the  thermal emission. 

Using the ISOPHOT 60 and 170\,$\mu$m data, \cite{Hippelein} found an anti-correlation between the flux density ratio of H$\alpha$/60\,$\mu$m and 170\,$\mu$m flux density, suggesting the extinction to be related to the cold dust in M33.  

We obtain the distribution of the dust optical depth at the H$\alpha$ wavelength  for the whole M33 (an extinction map) by analysis of dust emission and absorption using the high sensitivity and resolution Multiband Imaging Photometer Spitzer \citep[MIPS,][]{Rieke} FIR data at 70 and 160\,$\mu$m. This leads to an H$\alpha$ map corrected for extinction (de-extincted H$\alpha$ map), our free-free template.
The thermal and nonthermal maps at 3.6 and 20\,cm are obtained at an angular resolution of 90$\arcsec$ (equivalent to a linear resolution of 360\,pc) by both the new (using a free-free template) and the standard method (assuming a constant nonthermal spectral index) and the results are compared. Further, we determine variations of the nonthermal spectral index across M33 as detected by the new method and discuss the exponential scale lengths of both thermal and nonthermal emission. 

In Sect.~3 we derive distribution of dust color temperature. This temperature and the 160\,$\mu$m flux density are used to obtain the dust optical depth (extinction) and its distribution in Sects.~4 and 5.
We correct the H$\alpha$ emission for the extinction and then convert it to the thermal radio emission, following \cite{Dickinson} (Sect.~6). The nonthermal intensity and spectral index maps are produced in Sects.~7 and 8. We discuss and compare the results from the new and standard methods in Sect.~9. Finally, conclusions are presented in Sect.~10.

%\cite{Berkhuijsen_83} obtained the distribution of the thermal emission at 6.2\,cm from a catalogue of HII regions in H$\alpha$ \citep{Boulesteix} where diffuse emission was not included. Subtracting from the total radio 6.2\,cm emission, they also obtained distribution of the nonthermal emission. However, their inability to detect all the diffuse emission cause small scale lengths of both thermal and nonthermal components. 
%Decomposition of the radio continuum emission from M33 was first made by \cite{Berkhuijsen_83} at 6.2\,cm. They also used the H$\alpha$ emission as the thermal emission tempelate, but only considered the H$\alpha$ emission from the HII regions (using a catalogue of HII regions given by \citep{Boulesteix}), missing diffuse emission.  
%Using the standard separation method, \cite{Buczilowski_88} presented thermal and nonthermal maps at 6.3\,cm with  low angular resolution ($\sim$\,8$\arcmin$) and sensitivity. 

\section{Data}

The radio continuum data at 3.6 and 20\,cm are presented in Paper I \citep{Tabatabaei_2_07}. At 3.6\,cm, M33 was observed with the 100-m Effelsberg telescope of the MPIfR\footnote{The 100--m telescope at Effelsberg is operated by the Max-Planck-Institut f\"ur Radioastronomie (MPIfR) on behalf of the Max--Planck--Gesellschaft.}. The 20\,cm data were obtained from observations with the Very Large Array (VLA\footnote{The VLA is a facility of the National Radio Astronomy Observatory. The NRAO is operated by Associated Universities, Inc., under contract with the National Science Foundation.}) corrected for missing spacing using the Effelsberg data at 20\,cm.  
We also use the latest combined epoches of MIPS Spitzer data at 70 and 160\,$\mu$m \citep[as presented in ][]{Tabatabaei_1_07,Tabatabaei_1}. The H$\alpha$ map is from  Kitt Peak National Observatory (KPNO) \citep{Hoopes_et_al_97H}. Table~\ref{tab:data} summarizes the data used in this work.

\section{Temperature distribution of dust}

We derive the color temperature of the dust, $T$, between 70$\mu$m and 160$\mu$m
from the formula
\begin{equation}
\frac{F_1}{F_2}=\frac{\nu_1^{\beta}}{\nu_2^{\beta}} . \frac{B_1(T)}{B_2(T)} \ . 
\end{equation}

Here $F$ denotes the measured flux, $B(T)$ the Planck function, $\nu$ the
frequency, and 1 and 2 refer to the wavelength of 70$\mu$m and 160$\mu$m,
respectively.  The power law index of the absorption efficiency at FIR
wavelengths, $\beta$, is set to 2, which should be appropriate for interstellar
grains \citep{andriesse,Draine}.

\begin{figure}
\resizebox{7.3cm}{!}{\includegraphics*{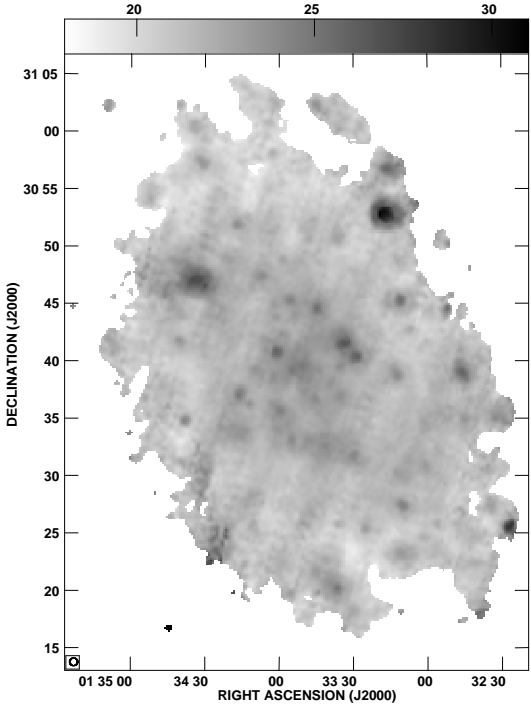}}
\caption[]{Dust temperature map of M33 obtained from I$_{70\mu m}$/I$_{160\mu m}$
ratio (only pixels with intensity above 3$\sigma$ level were used). The angular
resolution of 40$\arcsec$ is shown in the lower left-hand corner of the map. The
bar at the top gives the dust temperature in Kelvin.}
\label{fig:tempere}
\end{figure}

\begin{figure}
%\resizebox{7.3cm}{!}{\includegraphics*{m33-T.40arc.ps}}
\resizebox{7.3cm}{!}{\includegraphics*{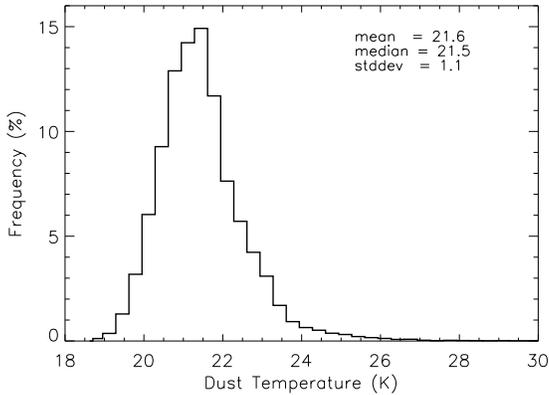}}
\caption[]{Histogram of the dust temperature map. It shows the population of
pixels as a function of the temperature intervals. The number of bins used for
this plot is 80.}
\label{fig:histtempere}
\end{figure}

The MIPS FIR maps at 70 and 160\,$\mu$m have been smoothed to an angular
resolution of 40$\arcsec$ and normalized to the same grid and center position.
The resulting dust temperature distribution is shown in Fig.~\ref{fig:tempere}, indicating
variations between 19 and 28\,K.
Fig.~\ref{fig:histtempere} displays the relative frequency of occurence of the
temperature intervals in 10$\arcsec \times 10\arcsec$ pixels.  The maximum is
attained at 21.5 K, close to the mean value of 21.6 K.  Warmer dust with
$T>25$\,K dominates in star-forming regions and in the center of the galaxy.
The highest temperatures are found in the HII complexes NGC\,604, NGC\,592, and
IC\,133.

\section{Optical depth at 160$\mu$m}

Having determined $T$, we obtain the optical depth from the equation
\begin{equation}
%I_{160\mu{\rm m}} = B_{160\mu {\rm m}}(T)\, (1 - \exp(-\tau_{160\mu {\rm m}})),
I_{160\mu{\rm m}} = B_{160\mu {\rm m}}(T)\, [1 - e^{-\tau_{160\mu {\rm m}}}] \ ,
\end{equation}
where $I_{160\mu{\rm m}}$ is the intensity.  The distribution of
$\tau_{160\mu{\rm m}}$ over the disk in M33 is plotted in Fig.~3.  Because the
temperature variations are very moderate, the optical depth $\tau_{160\mu {\rm
m}}$ usually follows the 160$\mu$m intensity map quite closely \citep[see
][]{Tabatabaei_1_07}.  Nevertheless, for a fixed intensity, a warm region (26\,K)
has a three times lower optical depth than a cold one  (19\,K).

\section{Distribution of extinction}

\begin{figure}
\resizebox{7.3cm}{!}{\includegraphics*{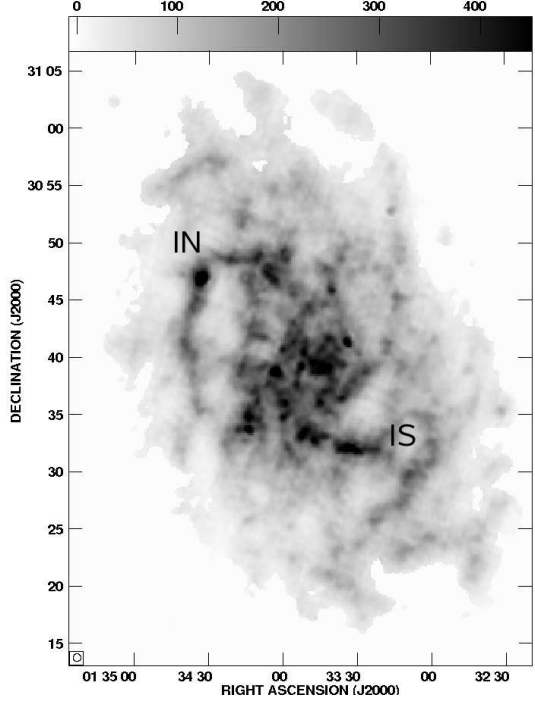}}
\caption[]{ Distribution of the dust optical depth at 160\,$\mu$m wavelength in
M33. The bar at the top shows $10^{6} \times \tau_{160\mu {\rm m}}$. The main
northern (IN) and southern (IS) spiral arms are indicated. The angular
resolution of 40$\arcsec$ is shown in the lower left-hand corner of the map.  }
\label{fig:tau}
\end{figure}

To convert $\tau_{160\mu {\rm m}}$ into the dust optical depth at the
wavelength of the H$\alpha$ line, $\tau_{{\rm H}\alpha}$, we have to multiply it 
by $\kappa_{{\rm H}\alpha}/ \kappa_{160\mu {\rm m}}$, the ratio of the
 dust extinction coefficient per unit mass at the corresponding wavelengths.
Using the extinction curve given by a standard dust model for the diffuse medium \citep[see, e.g. Figure 12.8 of][]{krugel}, we estimate $\tau_{{\rm H}{\alpha}} \simeq 2200\, \tau_{160\mu {\rm m}}$. Therefore, after multiplication with 2200,  Fig.~\ref{fig:tau} gives also the distribution of $\tau_{{\rm H}{\alpha}}$ across the galaxy at a linear resolution of 360\,pc.

$\tau_{{\rm H}{\alpha} }$ is around one half in the extended central region and
in the two main arms, IN and IS, and it is somewhat smaller in other arms
(mostly between 0.2 and 0.4).  In the center of the galaxy and in massive star
forming regions, specifically in the southern arm IS and the HII complexes
NGC604, NGC595, and B690, $\tau_{{\rm H}{\alpha} }$ exceeds 0.7.  The highest
dust optical depth one finds in the center of the galaxy and in NGC604 where
$\tau_{{\rm H}\alpha} \simeq 0.97$ and 0.88, respectively (at the linear
resolution of 360\,pc).  Therefore, M33 is generally almost transparent for
photons with $\lambda \simeq 6560$\AA\ propagating towards us.

To estimate how much the detected H$\alpha$ radiation has been attenuated, we
note that H$\alpha$ photons are usually emitted from sources {\it within} the
galaxy.  The optical depth $\tau_{{\rm H}\alpha}$ therefore only gives an upper
limit.  Following \cite{Dickinson}, we set the effective thickness to $\tau_{\rm
eff} = f_d \times \tau_{{\rm H}\alpha}$ with $f_d < 1$; the attenuation factor
for the H$\alpha$ flux is then $e^{-\tau_{\rm eff}}$.

At 360\,pc resolution, one may assume that the H$\alpha$ emitters, ionized gas
in HII regions and diffuse gas, are uniformly mixed with the dust, which would
imply $f_d \simeq 0.5$.  \cite{Dickinson} found $f_d= 0.33$ for the Milky Way
(because the z-distribution of the H$\alpha$ emission is smaller than that of
the dust) and we adopt their value also for M33.  But, as will be shown in
Sect. 9.4, the determination of the thermal fraction of the radio emission is
not very sensitive to the particular choice of $f_d$.

Of course, it would be preferable not to use a uniform value $f_d$ for the whole
galaxy, but one that is adapted to the geometry \citep[well mixed diffuse medium
or shell-like in HII regions, ][]{Witt} and the dust column density.
This needs to specify the location of the stellar sources and the absorbing dust and to solve the radiative transfer problem with massive numerical computations. 

%\cite{Gordon_00} developed a flux ratio method to determine the dust attenuation which accounts for different geometries using a Monte Carlo radiative transfer model together with a stellar evolutionary synthesis model. These models are not used in this study, as they would require additional assumptions as well as more data to constrain them.

\begin{figure}
\resizebox{7.3cm}{!}{\includegraphics*{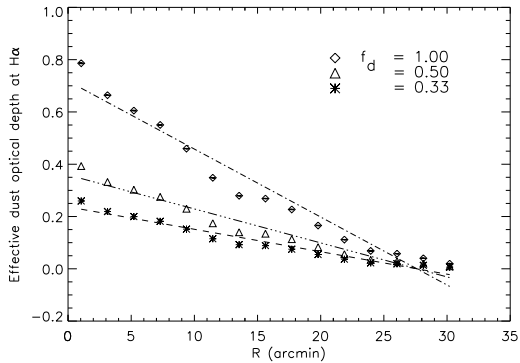}}
\caption[]{Radial distribution of the mean effective optical depth at H$\alpha$
($\tau_{\rm eff}$) in rings of width of 0.5\,kpc in the galactic plane for
$f_d$\,=\,0.33, 0.5 and 1.0. The errors are smaller than the size of the
symbols. }
\label{fig:tauscatter}
\end{figure}

An interesting related question is how the extinction in M33 changes with
galactic radius $R$. To investigate this, we integrated Spitzer FIR
flux densities at 70 and 160\,$\mu$m in rings of 0.5\,kpc width in the galactic
plane \citep[inclination of 56$^{\circ}$, ][]{Regan_etal_94}.  We first derived
the mean dust temperature of each ring and then the mean dust optical depth
$\tau_{\rm eff}$.  Fig.~\ref{fig:tauscatter} shows $\tau_{\rm eff}$ versus the
galactocentric radius $R$.  For $f_d=0.33$, the dependence can be described by
$\tau_{{\rm H} _{\alpha}} ({\rm R})= (-0.009 \pm 0.002)\,{\rm R} + (0.24 \pm
0.03)$.  As $f_d$ rises, the slope gets steeper.  The case $f_d=1$ (for the full
layer), which is appropriate for pure background sources of M33, is shown for
comparison. Generally, a decrease of the extinction with galactocentric radius is expected as it
reflects the decrease in surface density towards the periphery. 

From a comparison of H$\alpha$ data with radio flux densities at 6.3 and 21\,cm, \cite{Israel_80} and \cite{Berkhuijsen_82,Berkhuijsen_83} found a rather steep radial gradient of the extinction:
considering 8 bright HII regions, \cite{Berkhuijsen_82} derived the relation
$\tau_{{\rm H}\alpha} = A_{{\rm H}\alpha}/1.086 = (-0.038 \pm 0.003)\,{\rm
R}\,+\,(1.45\pm 0.05)$, with $R$ in arcmin.  Qualitatively similar results were
reported by \cite{Peterson_97}, using Balmer and Paschen emission lines, and by
\cite{Hippelein} based on a study of the H$\alpha$ to the 60\,$\mu$m ISOPHOT
flux ratio in HII regions.  The only discrepant result by
\cite{Devereux_etal_97} who, from their pixel-to-pixel comparison of H$\alpha$
and 6\,cm thermal radio maps, did not find any systematic decline in $A_{\rm
V}$, \cite{Hippelein} were able to explain by assuming that the radial gradient
of the cold dust (170\,$\mu$m ISOPHOT) is stronger in the HII regions than in
the diffuse gas.

\begin{figure*}
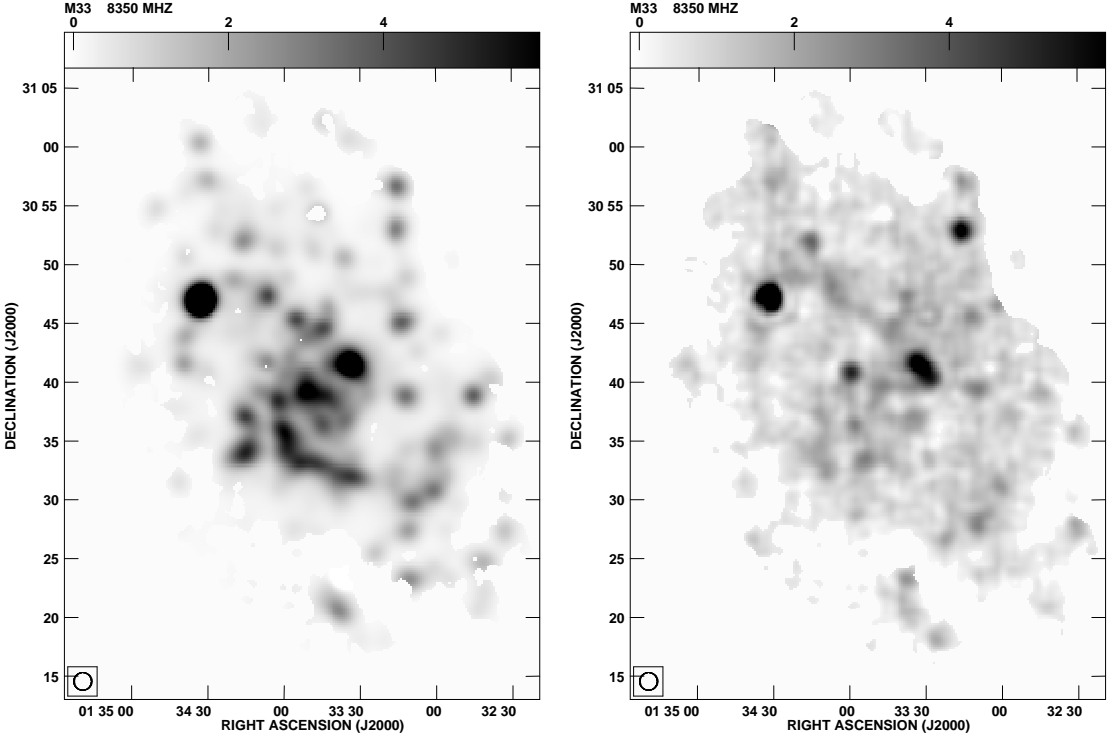

\resizebox{15cm}{!}{\includegraphics*{3cm-th.ps}
\includegraphics*{m33-nth.ps}}
\caption[]{Thermal ({\it left}) and nonthermal ({\it right}) maps at
3.6\,cm. The grey-scale gives the flux density in $\mu$Jy/beam. The angular
resolution is 90$\arcsec$ (shown in the lower left of the images) with a grid
size of 10$\arcsec$. The bright HII complexes NGC604 (RA\,=\,1$^h$ 34$^m$
32.9$^s$ $\&$ DEC\,=\,30$^{\circ}$ 47$\arcmin$ 19.6$\arcsec$) and NGC 595
(RA\,=\,1$^h$ 33$^m$ 32.4$^s$ $\&$ DEC\,=\,30$^{\circ}$ 41$\arcmin$
50.0$\arcsec$) are clearly visible in both maps, whereas IC133 (RA\,=\,1$^h$
33$^m$ 15.3$^s$ and DEC\,=\,30$^{\circ}$ 53$\arcmin$ 19.7$\arcsec$) appears
stronger in the nonthermal map.  }
\label{fig:ther3.6cm}
\end{figure*}

\begin{figure*}
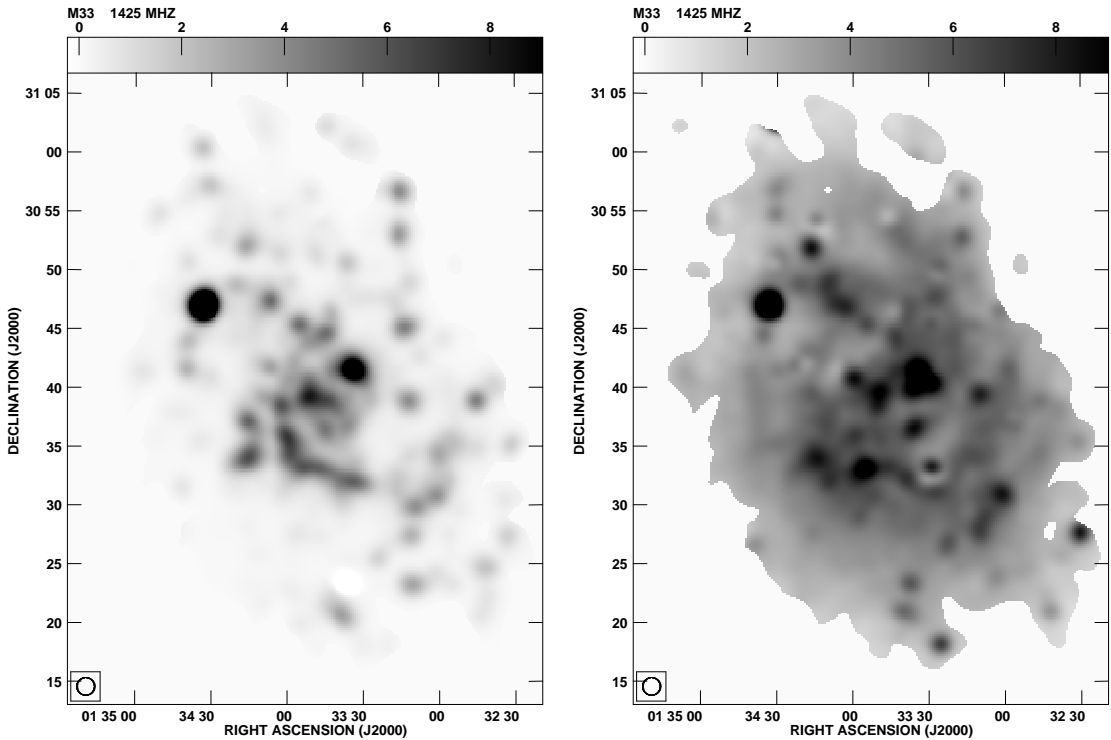

\resizebox{15cm}{!}{\includegraphics*{20cm-th.ps}
\includegraphics*{20cm-nth.ps}}
\caption[]{Thermal ({\it left}) and nonthermal ({\it right}) maps at
20\,cm. The grey-scale gives the flux density in $\mu$Jy/beam. The angular
resolution is 90$\arcsec$ (shown in the lower left of the images) with a grid
size of 10$\arcsec$. }
\label{fig:ther20cm}
\end{figure*}

\section{Determination and distribution of free-free emission}

From the observed H$\alpha$ intensity, $I$, and the effective extinction,
$\tau_{\rm eff}$, we derive the intrinsic H$\alpha$ intensity, $I_0$, according
to

\begin{equation}
I = I_0 \,\,e^{-\tau_{\rm eff}} \ .
\end{equation}

%One might expect that the brighter HII regions need more correction for
%extinction. However, the decreasing effect of the dust temperature on the
%optical depth should also be taken into account (Eq. (2)). A good example is
%IC133, a giant HII region which has little extinction as it is the warmest
%region (see Fig.~\ref{fig:tau}).

Integration of the H$\alpha$ maps out to a radius of 7.5\,kpc yields a ratio of
corrected to observed total H$\alpha$ flux density of 1.13 for $f_d=0.33$ and
1.25 for $f_d=0.5$.  Thus only 13$\%$ (25$\%$) of the total H$\alpha$ emission
is obscured by dust. The small extinction within M33 was predicted by the
wavelet study of 3.6\,cm and H$\alpha$ emission \citep{Tabatabaei_1_07}.
%\cite{Dickinson} showed that the H$\alpha$ emitting medium in our Galaxy is
%optically thick to ionizing Lyman photons \citep[case B, ][]{Osterbrock} not
%only for HII regions ($\tau \sim 10^3 - 10^{10}$) but also for faint H$\alpha$
%features at intermediate and high Galactic latitudes ($\tau \sim 1 - 30$). In
%the case of M33, we receive H$\alpha$ emission from a combination of HII 
%regions and diffuse ionized medium along the line of sight. Therefore, the
%average thickness of the ionized medium to the Lyman continuum attributed to
%each pixel of the H$\alpha$ image would be a value between the minimum and
%maximum optical depths already mentioned, $ 1<\tau <10^{10}$. Thus, we may also
%consider case B for M33 through the \cite{Valls} expression for the H$\alpha$
%intensity (in units of erg\,cm$^{-2}$\,s$^{-1}$ sr$^{-1}$)\,:

The emission measure (EM) follows from the H$\alpha$ intensity via the
expression \cite{Valls}

\begin{equation}
I_{{\rm H}\alpha} = 9.41 \times 10^{-8} T^{-1.017}_{e4}
10^{-\frac{0.029}{T_{e4}}} \, {\rm EM} \ ,
\end{equation}
where the electron temperature, $T_{e4}$, is in units of $10^4$\,K, EM in
cm$^{-6}$ pc, and it is assumed that the optical depth of HI resonance lines is
large (usually denoted as case B).  The emission measure is
related to the continuum optical thickness, $\tau_c$, of the ionized gas by
\begin{equation}
\tau_c = 8.235 \times 10^{-2} a T_e^{-1.35} \nu_{{\rm GHz}}^{-2.1} (1+0.08) \,
{\rm EM} \ ,
\end{equation}
with $a \simeq 1$ \citep{Dickinson}.  The factor (1\,+\,0.08) takes into account the contribution from the  singly ionized He. The brightness temperature of the radio
continuum emission, $T_b$, then follows from
\begin{equation}
T_b = T_e(1-e^{-\tau_c}) \ .
\end{equation}

\begin{table}
%\begin{center}
\caption{Thermal flux density and thermal fraction at 3.6\,cm ($f_d$\,=\,0.33 and $T_e$\,=\,10000\,k).}
\begin{tabular}{ l l l l } 
\hline \hline
Object &  Observed  & Thermal  & Thermal \\
     &  flux density & flux density& fraction \\
 & (mJy)& (mJy) & $\%$ \\
\hline
NGC604 &  50.44\,$\pm$\,0.13 & 38.55\,$\pm$\,0.08  & 76.4\,$\pm$\,0.3\\
NGC595 & 17.95\,$\pm$\,0.18 & 12.13\,$\pm$\,0.11   & 67.6\,$\pm$\,1.3\\
IC133 &  9.08\,$\pm$\,0.11 & 2.66\,$\pm$\,0.04  & 29.3\,$\pm$\,0.8\\
B690 &  4.51\,$\pm$\,0.11 &  3.11\,$\pm$\,0.10 & 68.9\,$\pm$\,3.9\\
B61/62 & 3.72\,$\pm$\,0.11 & 2.24\,$\pm$\,0.07  & 60.2\,$\pm$\,3.7\\
IC132 &  4.89\,$\pm$\,0.05 & 3.25\,$\pm$\,0.02 & 66.5\,$\pm$\,1.0\\
IC131 &  4.59\,$\pm$\,0.07 &  3.26\,$\pm$\,0.05 & 71.0\,$\pm$\,2.2 \\
NGC588 &  4.64\,$\pm$\,0.09 & 3.24\,$\pm$\,0.06  & 69.8\,$\pm$\,2.6\\
IC142 &  3.51\,$\pm$\,0.08 & 2.73\,$\pm$\,0.10  & 77.8\,$\pm$\,4.6\\
B691 &  4.60\,$\pm$\,0.13 &  2.35\,$\pm$\,0.06 & 51.1\,$\pm$\,2.6\\
NGC592 &  4.28\,$\pm$\,0.08 & 2.76\,$\pm$\,0.06  & 64.5\,$\pm$\,2.6\\\hline
M33 (mJy): & &&\\
R$<7.5$\,kpc & 761\,$\pm$\,63 & 391\,$\pm$\,74 & 51.4\,$\pm$\,4.2  \\
%(mJy) & &  &\\
\hline
\end{tabular}
%\end{center}
\label{tab:3.6new}
\end{table}
\begin{table}
%\begin{center}
\caption{Thermal flux density and thermal fraction at 20\,cm ($f_d$\,=\,0.33 and $T_e$\,=\,10000\,k).}
\begin{tabular}{ l l l l } 
\hline \hline
Object &  Observed  & Thermal  & Thermal \\
     &  flux density & flux density& fraction \\
 & (mJy)& (mJy) & $\%$ \\
\hline
NGC604 &  62.75\,$\pm$\,0.21 & 45.46\,$\pm$\,0.09  & 72.4\,$\pm$\,0.4\\
NGC595 & 20.50\,$\pm$\,0.22 & 14.31\,$\pm$\,0.14   & 69.8\,$\pm$\,1.4\\
IC133 &  5.35\,$\pm$\,0.07 & 3.11\,$\pm$\,0.03  & 58.1\,$\pm$\,0.9\\
B690 &  5.80\,$\pm$\,0.12 &  3.69\,$\pm$\,0.12 & 63.6\,$\pm$\,2.4\\
B61/62 & 4.00\,$\pm$\,0.21 & 2.69\,$\pm$\,0.08  & 67.2\,$\pm$\,4.0\\
IC132 &  5.52\,$\pm$\,0.05 & 3.81\,$\pm$\,0.02 & 69.0\,$\pm$\,0.7\\
IC131 &  4.12\,$\pm$\,0.07 &  3.84\,$\pm$\,0.06 & 93.2\,$\pm$\,2.1 \\
NGC588 &  4.67\,$\pm$\,0.17 & 3.74\,$\pm$\,0.06  & 80.0\,$\pm$\,3.2\\
IC142 &  4.14\,$\pm$\,0.08 & 3.09\,$\pm$\,0.13  & 74.6\,$\pm$\,3.4\\
B691 &  7.40\,$\pm$\,0.20 &  2.43\,$\pm$\,0.04 & 32.8\,$\pm$\,1.0\\
NGC592 &  5.43\,$\pm$\,0.14 & 3.27\,$\pm$\,0.08  & 60.2\,$\pm$\,2.1\\ \hline
M33 (mJy): & &&\\
R$<7.5$\,kpc & 2722\,$\pm$\,60 & 478\,$\pm$\,85 & 17.6\,$\pm$\,0.4 \\
%(mJy) & & & \\
\hline
\end{tabular}
%\end{center}
\label{tab:20new}
\end{table}

Whereas the electron temperature in the Milky Way is known to increase with
galactocentric radius \citep{Shaver} as a result of decreasing metallicity
\citep{Panagia}, M33 does not show significant variations in metallicity
\citep[e.g. ][]{Willner,Margini}.  \cite{Crockett} derived the electron
temperature from forbidden line ratios in 11 HII regions with galactocentric
distances from 1 to 7\,kpc.  Their $T_e$ values range from 7300\,K to 12800\,K
with a mean value of 10000\,K and no clear radial gradient.  As there are no
$T_e$ measurements for the diffuse ionized gas in M33, we adopt a fixed value of
$T_e=10000$\,K.

The conversion factors from brightness temperature (K) to the observed radio
flux density (Jy/beam) are 2.6 and 74.5 at 3.6 and 20\,cm, respectively.  The
final free-free maps are shown in Figs.~\ref{fig:ther3.6cm} and
\ref{fig:ther20cm}.  At both wavelengths, the strongest thermal emission emerges
from HII regions, in particular, the HII complexes NGC604 and NGC595, but the
southern arm IS (Paper~I) and the center of the galaxy are also very bright.  At
both wavelengths, the thermal fraction of the diffuse emission is $<$\,25$\%$ in
interarm regions. The average error in the thermal fraction obtained using the
error propagation method is 7$\%$.
% with a standard deviation of $4\%$. 

Integrating the thermal maps in rings around the galaxy center out to a radius
of 7.5\,kpc, we obtain the total thermal flux densities and thermal fractions at
3.6 and 20\,cm (see Table~\ref{tab:3.6new} and \ref{tab:20new}). These Tables
also give the thermal flux densities and thermal fractions at the position of
the 11 brightest HII regions (the coordinates of these HII complexes are listed
in Paper~I).

\section{Distribution of nonthermal emission}

Subtracting the maps of the thermal emission from the observed maps at each wavelength, maps of the nonthermal  emission are obtained (Figs.~\ref{fig:ther3.6cm} and \ref{fig:ther20cm}). The latter maps exhibit diffuse emission extending to large radii. They also show strong features in the spiral arms and central region of the galaxy. The strongest nonthermal emission emerges from the HII complexes NGC604 and NGC595 at both 3.6 and 20\,cm. Typically, HII complexes host tens of young O/B stars, many of which end as supernovae whose remnants contribute to a mixed (thermal and nonthermal), flat spectrum of the total radio emission from these regions \citep[as discovered in NGC\,604 and NGC\,595,  ][]{Dodorico_78,Gordons_93,Yang_96}.  Supernova remnants with central energy sources in the form of young pulsars (Crab-like remnants) also have flat nonthermal spectra. 

Another strong feature in the 20\,cm nonthermal map belongs to the active star-forming region in the central southern arm IS. This confirms our previous conclusion about the spatial coupling of the nonthermal emission with the starforming regions (Paper~I). There are also other point-like sources which are not resolved.  Looking at the 3.6\,cm map, the giant HII region IC133 (RA\,=\,1$^h$ 33$^m$ 15.3$^s$ and DEC\,=\,30$^{\circ}$ 53$\arcmin$ 19.7$\arcsec$) is also strong, but not at 20\,cm. Hosting the two strongest optically thick HII regions \citep{Johnson}, IC133 has an inverted spectrum. The appearance of this source in the 3.6\,cm nonthermal map is due to the assumption that the free-free emission is optically thin (equivalent to the thermal spectral index of 0.1 when S$_{th} \sim \nu^{-0.1}$). Furthermore, IC133 probably contains some nonthermal emission as \cite{Schulman} found it associated with a bright X-ray source located in a hole in the HI layer of the galaxy, indicating energetic stellar winds and supernovae from massive stars.

\begin{figure}
%\begin{center}
\resizebox{7.3cm}{!}{\includegraphics*{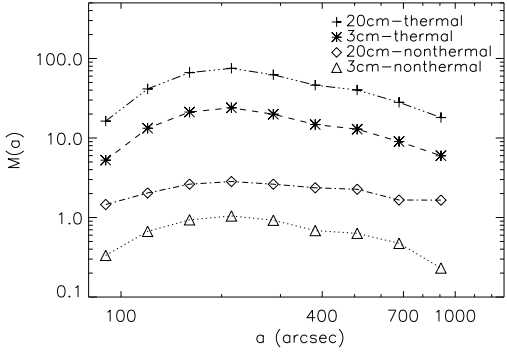}}
\caption[]{The wavelet spectra of the 3.6 and 20\,cm thermal and nonthermal
emission at $90\arcsec$ resolution. The data points correspond to the scales 90,
120, 160, 214, 286, 381, 509, 679, 907$\arcsec$. The spectra are shown in
arbitrary units. 100$\arcsec$ corresponds to 400\,pc in M33.}
%\end{center}
\label{fig:waveletspectra}
\end{figure}

\begin{figure}
\resizebox{7.5cm}{!}{\includegraphics*{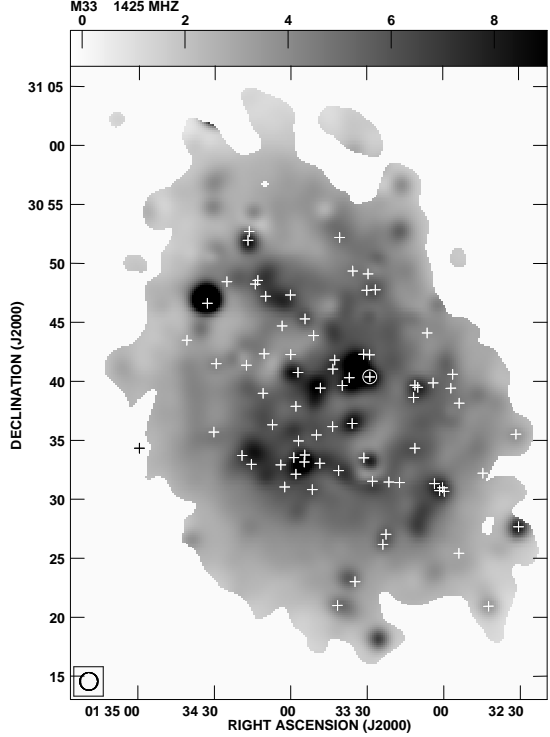}}
\caption[]{The same 20\,cm nonthermal map as shown in Fig.~\ref{fig:ther20cm}
with supernova remnants (crosses) superimposed. The supernova remnants with a signal-to-noise ratio of
larger than 0.7 were selected from the catalogue of \cite{Gordons_98}. The
circle shows an HII region with nonthermal spectral index \citep{Gordons_99}. }
\label{fig:20cmsnr}
\end{figure}

\cite{Tabatabaei_1_07} discussed the wavelet spectrum of the total radio continuum maps. Here, we present the wavelet energy of the thermal and nonthermal emission at different scales. Fig.~\ref{fig:waveletspectra} shows that the wavelet spectrum of the nonthermal emission is smoother than that of the thermal emission at both wavelengths. Furthermore, the distribution of the nonthermal wavelet energy is smoother at 20\,cm than at 3.6\,cm. Scales smaller than the width of the spiral arms ($\simeq 400\arcsec$ or 1.6\,kpc) and larger than the size of the giant starforming regions ($\simeq 100\arcsec$ or 0.4\,kpc) at 20\,cm are not as prominent as those at 3.6\,cm, where the synchrotron photons have higher energies. In fact the synchrotron emission distribution reflects the combined distribution of the interstellar magnetic field and the cosmic ray electrons. Hence, assuming that the distribution of the interstellar magnetic field is the same at both wavelengths, one can conclude: the higher the energy of CRs, the more dominant the localized structures (which can be linked to the cosmic ray `sources' e.g. supernova remnants). This is expected as energy losses of CR electrons increase with electron energy and field strength. Studying propagation effects with self-consistent Galactic wind simulations, \cite{Breitschwerdt} predicted a similar energy-dependence for the CR nucleon distribution in our Galaxy. 
They expect that the high energy (TeV) $\gamma$-rays from CR sources dominate the diffuse $\gamma$-ray emission, while the Galactic $\gamma$-ray observations in the GeV range (with EGRET and COS-B) have shown a roughly uniform distribution of the $\gamma$-ray emissivity in the Galactic plane.  

The dominant scales of the wavelet spectra of the thermal (at 3.6 and 20\,cm) and the 3.6\,cm nonthermal emission are the same, indicating that the distribution of the cosmic ray sources is similar to that of the thermal sources. At scales larger than  a\,=\,$214\,\arcsec$, the 3.6\,cm nonthermal spectrum decreases twice as fast as the 20\,cm nonthermal spectrum. The smaller energy losses of the CR electrons with lower energies (at 20\,cm) than of those with higher energies (at 3.6\,cm) may cause a smoother distribution of the emission at all scales. This indicates that besides the CR sources embedded in the star-forming regions, there is another component of the synchrotron emission in the form of a diffuse disk or a galactic halo that is better visible at 20\,cm than 3.6\,cm. This could be verified by multiwavelength observations of edge-on galaxies at resolutions higher than those of the existing data. 
\begin{figure}
\resizebox{7.3cm}{!}{\includegraphics*{ntspect.cont.ps}}
\resizebox{7.3cm}{!}{\includegraphics*{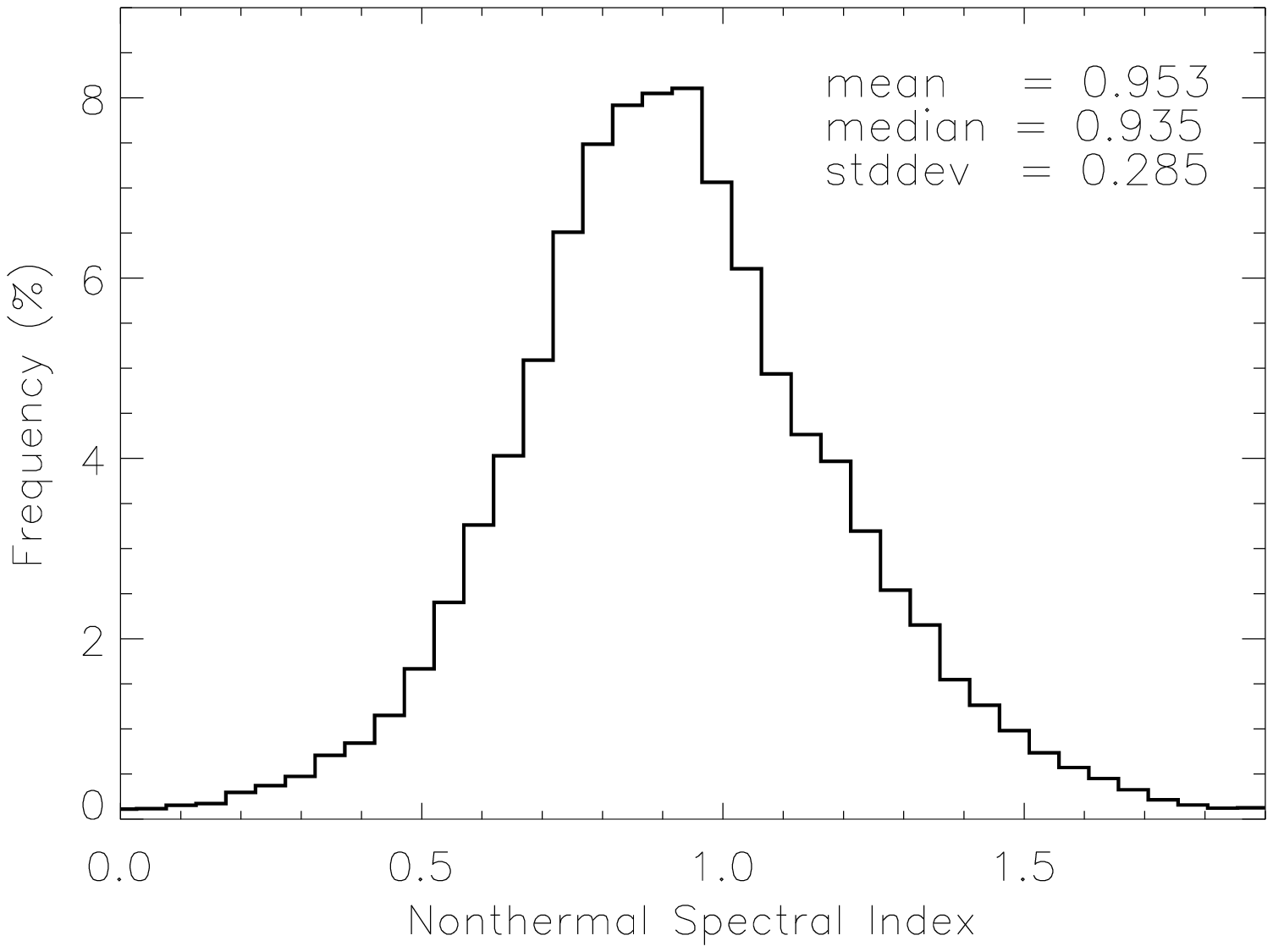}}
\caption[]{Nonthermal spectral index ($\alpha_{n}$) map obtained from the nonthermal radio fluxes at 3.6 and 20\,cm. The spiral arms are indicated by contours of the total radio emission at 3.6\,cm superimposed. Contour levels are 1.5, 4.5, and 12\,mJy/beam. The angular resolution is 90$\arcsec$ with a grid size of 10$\arcsec$. {\it Bottom:} A histogram of the nonthermal spectral index map.  The number of bins used for this plot is 50. }
\label{fig:ntspect}
\end{figure}

Fig.~\ref{fig:20cmsnr} shows the  the supernova remnants from \cite{Gordons_98} superimposed on the nonthermal 20\,cm map. There is general coincidence between nonthermal features and supernova remnants, the powerful sources of the relativistic electrons. \cite{Gordons_99} diagnosed 17 supernova remnants embedded in HII regions. They also found some 30 HII regions with nonthermal radio components of which it was not clear whether they belong to these regions or were external radio sources. The circle in Fig.~\ref{fig:20cmsnr} shows the position of the strongest  (S$_{20}$\,=\,7.0\,$\pm$\,0.2\,mJy, $\alpha \simeq 0.2$) HII region with nonthermal emission. As this HII region is not resolved in our thermal maps with 90$\arcsec$ resolution, it is not listed amongst the bright HII regions in Tables~2 and 3.

\section{Nonthermal spectral index}

From the nonthermal radio fluxes at 3.6 and 20\,cm, we obtained the spectral index of the nonthermal emission which was only computed for pixels with flux densities of at least three times the rms noise $\sigma$ at both frequencies. Fig.~\ref{fig:ntspect} shows that the nonthermal spectral index, $\alpha_n$, has a clumpy distribution. Note that the fainter regions have steeper spectra. The most probable value of $\alpha_n$ distributed across the galaxy is 0.95 (see the lower panel in Fig.~\ref{fig:ntspect}). 
%The average error in the nonthermal spectral index map is 24$\%$ with a standard deviation of 8$\%$ from error propagation method. 
In the star-forming regions, the nonthermal spectrum is relatively flat with an average value of $\alpha_n$ of 0.6$\pm$0.1, the typical spectral index of supernova remnants, but $\alpha_n$ increases to $1.2 \pm 0.2$ in the interarm regions and outer parts of the galaxy. 
This indicates energy losses of the relativistic electrons while they diffuse away from their origin in star-forming regions towards the interarm regions and the outer parts of the galaxy. For the first time, a nonthermal spectral index map can be used to achieve more realistic models for the propagation of CR electrons.
% that could also be applicable for the Milky Way where there is no diagnosis of the arms and interarms. 

%
\begin{figure}
%\begin{center}
\resizebox{7.3cm}{!}{\includegraphics*{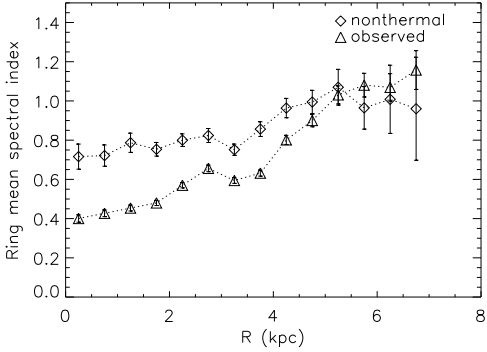}}
\caption[]{ Mean nonthermal spectral index in rings of 0.5\,kpc width in the galactic plane versus galactocentric radius. The total (observed) spectral index is also shown for comparison. The minima at $3<$R$<4$\,kpc is due to the HII complex NGC604.  }
%\end{center}Only data points with flux density larger than 3$\sigma$ level were considered. 
\label{fig:ringntsp}
\end{figure}

Integration of the nonthermal intensity in rings of 0.5\,kpc in the galactic plane yields the mean nonthermal spectral index in each ring. The radial variation of the ring mean spectral index is shown in Fig.~\ref{fig:ringntsp}, where the ring mean spectral index of the total radio continuum emission is also plotted for comparison. 
Up to R\,$\simeq$\, 4\,kpc, there is no increasing trend for the nonthermal spectral index with radius as for the total spectral index. Here the ring mean nonthermal spectral index varies between 0.65 and 0.9, indicating that CRs are injected by sources related to massive stars, and under leakage loss \citep{Biermann_95}.
Towards the outer parts of the galaxy the nonthermal and total spectral indices converge, confirming that the total radio continuum emission is mostly nonthermal at R\,$>4.5$\,kpc. This region corresponds to the synchrotron and inverse Compton loss dominated regime as $\alpha_n$ $\simeq 1$ on the average \citep{Biermann_95}.

\section{Discussion}

\subsection{Comparison with the standard method}

In this section, we first obtain the distribution of the thermal and nonthermal emission assuming that the nonthermal spectral index is constant across the galaxy (standard method). Then we compare the results from the two methods. 

\begin{figure*}
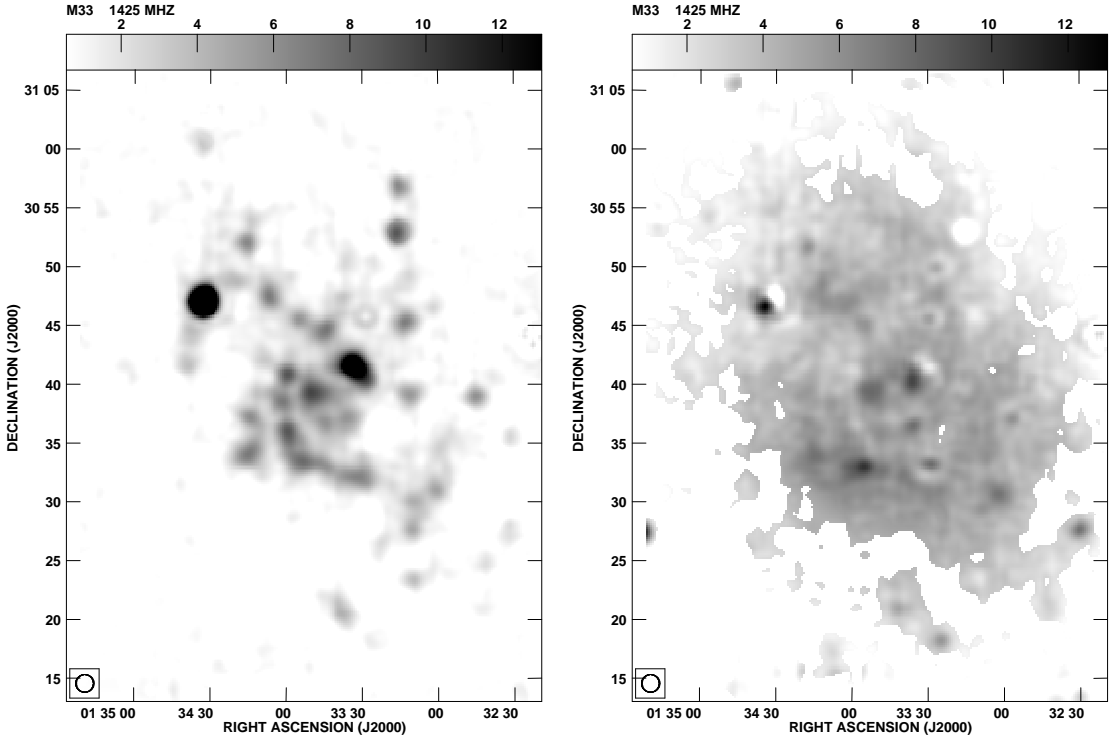

\resizebox{15cm}{!}{\includegraphics*{20cm-oldth.ps}
\includegraphics*{20cm-oldnth.ps}}
\caption[]{Thermal and nonthermal 20\,cm maps obtained from the standard method. The grey-scale gives the flux density in $\mu$Jy/beam. The angular resolution is 90$\arcsec$ with a grid size of 10$\arcsec$. }
\label{fig:oldth20cm}
\end{figure*}

Because the thermal emission is weak in the outer parts of the galaxy, one may consider the total spectral index from these parts as the pure nonthermal spectral index\footnote{This is confirmed by the new method, as mentioned in Sect.~8.}, $\alpha_n$ \citep[e.g. ][]{Berkhuijsen_03}. The total spectral index map of M33 gives $\alpha_n=1.0 \pm 0.1$ \citep[see Fig.~11, ][]{Tabatabaei_2_07}. For a total spectral index, $\alpha$, obtained from the observed flux densities at frequencies $\nu_1$ and $\nu_2$ and the constant value of $\alpha_n$, the thermal fraction at frequency $\nu_1$ is given by 

\begin{equation}
F_{th}^{\nu_1}=((\frac{\nu_2}{\nu_1})^{-\alpha}-(\frac{\nu_2}{\nu_1})^{-\alpha_n})/((\frac{\nu_2}{\nu_1})^{-0.1}-(\frac{\nu_2}{\nu_1})^{-\alpha_n}),
\end{equation}
\citep{Klein_84}. Then the thermal flux density at frequency $\nu_1$, $S_{th}^{\nu_1}$, is obtained from

\begin{equation}
S_{th}^{\nu_1}=S^{\nu_1} \times F_{th}^{\nu_1},
\end{equation}
and the nonthermal flux density at frequency $\nu_1$, $S_{n}^{\nu_1}$ is
  
\begin{equation}
S_{n}^{\nu_1}=S^{\nu_1} - S_{th}^{\nu_1}.
\end{equation}
Using the data at 3.6 and 20\,cm in the above formulae, the corresponding thermal and nonthermal maps are derived; those at 20\,cm are shown in Fig.~\ref{fig:oldth20cm}. The main difference between the standard and our new method (Fig.~\ref{fig:ther20cm}) concerns the distribution of the nonthermal emission, while the thermal maps show almost the same structures. The nonthermal emission from the standard method is weaker than that from the new method and hardly shows emission from star-forming regions in the arms. Often it is even weaker in the arms than in between the arms and in the outer parts of the galaxy.

In contrast to what is assumed in the standard method, large variations of the nonthermal spectral index are found across M33 by the new method (Fig. \ref{fig:ntspect}). We interpret this as clear evidence that CR  electrons suffer energy losses diffusing away from their places of origin in the arms towards interarm and outer regions. 

%\begin{figure}
%\resizebox{7.3cm}{!}{\includegraphics*{3cm-oldth.ps}}
%\caption[]{Thermal 3.6\,cm map obtained from the standard method. The angular resolution is 90$\arcsec$ with a grid size of 10$\arcsec$. }
%\label{fig:oldth3cm}
%\end{figure}

%\begin{figure}
%\resizebox{7.3cm}{!}{\includegraphics*{3cm-oldnth.ps}}
%\caption[]{Nonthermal 3.6\,cm map obtained from the standard method. The angular resolution is 90$\arcsec$ with a grid size of 10$\arcsec$. }
%\label{fig:oldnth3cm}
%\end{figure}
%

\begin{figure*}
\resizebox{15cm}{!}{\includegraphics*{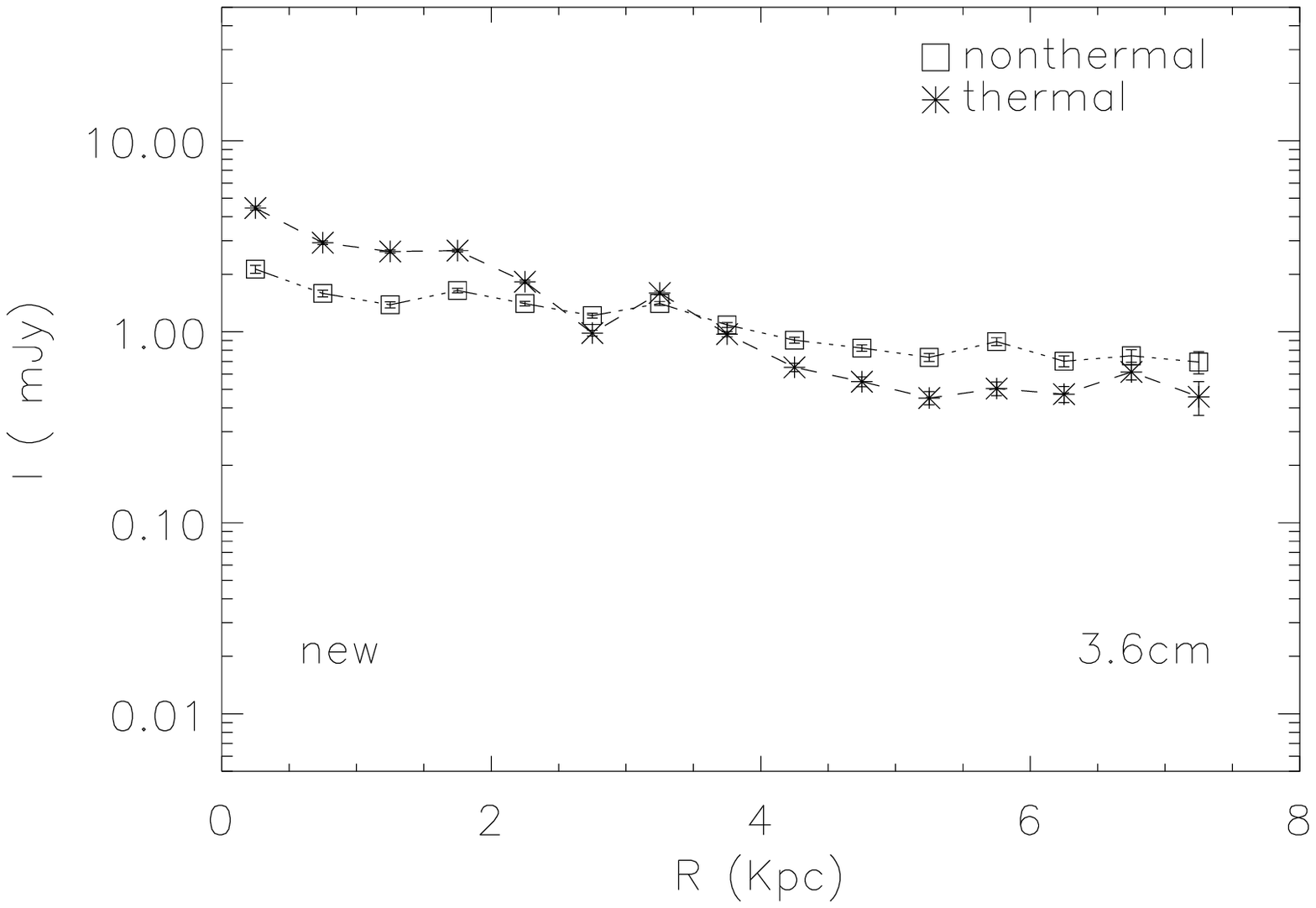}
\includegraphics*{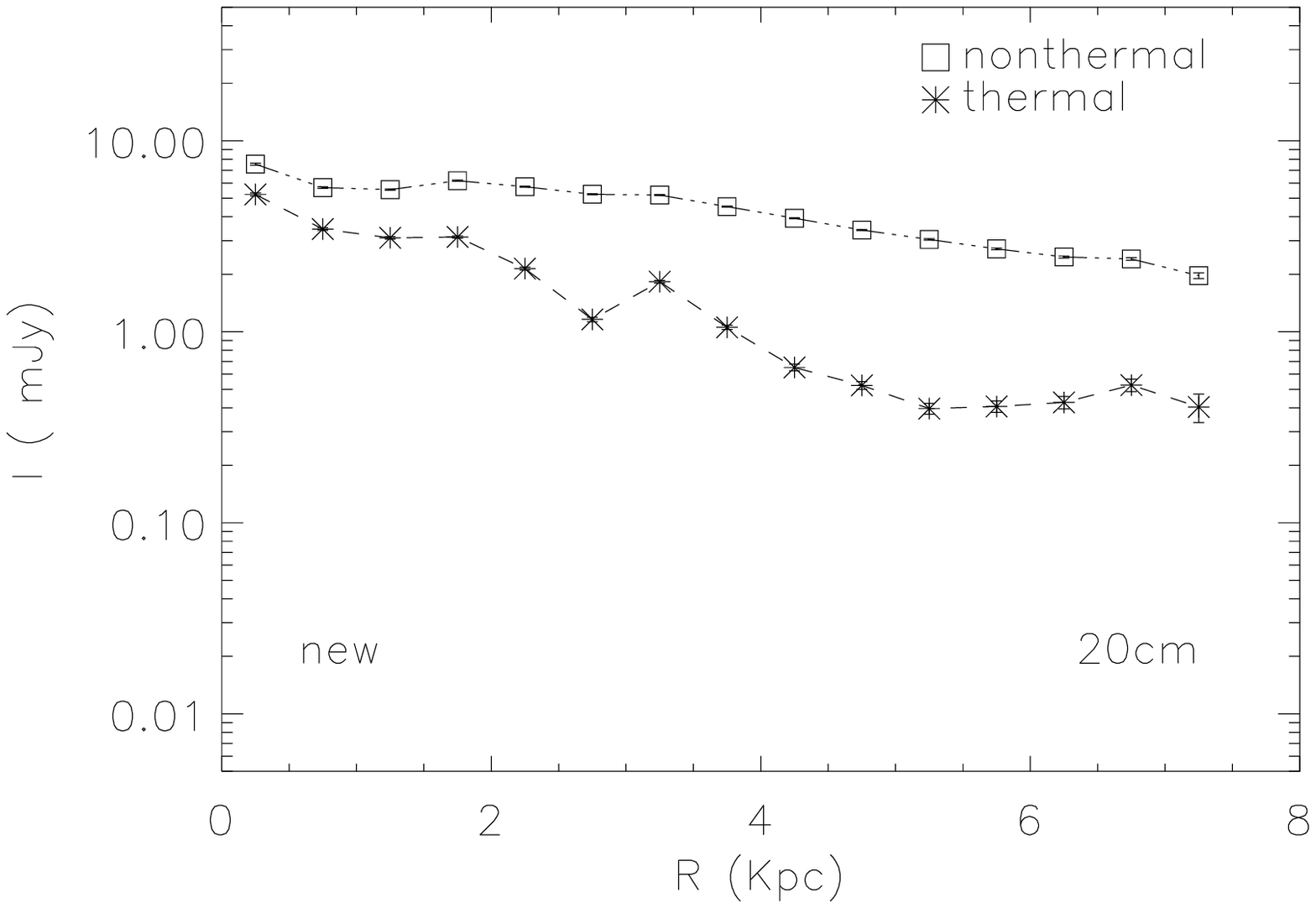}}
\caption[]{Radial profiles of the flux densities of the thermal and nonthermal emission at 3\,cm (left) and 20\,cm (right) obtained from the {\it new} separation method. The thermal flux densities show a bump at $3<$R$<4$\,kpc which is due to the HII complex NGC604.}
\label{fig:surf3cm}
\end{figure*}

\begin{figure*}
\resizebox{15cm}{!}{\includegraphics*{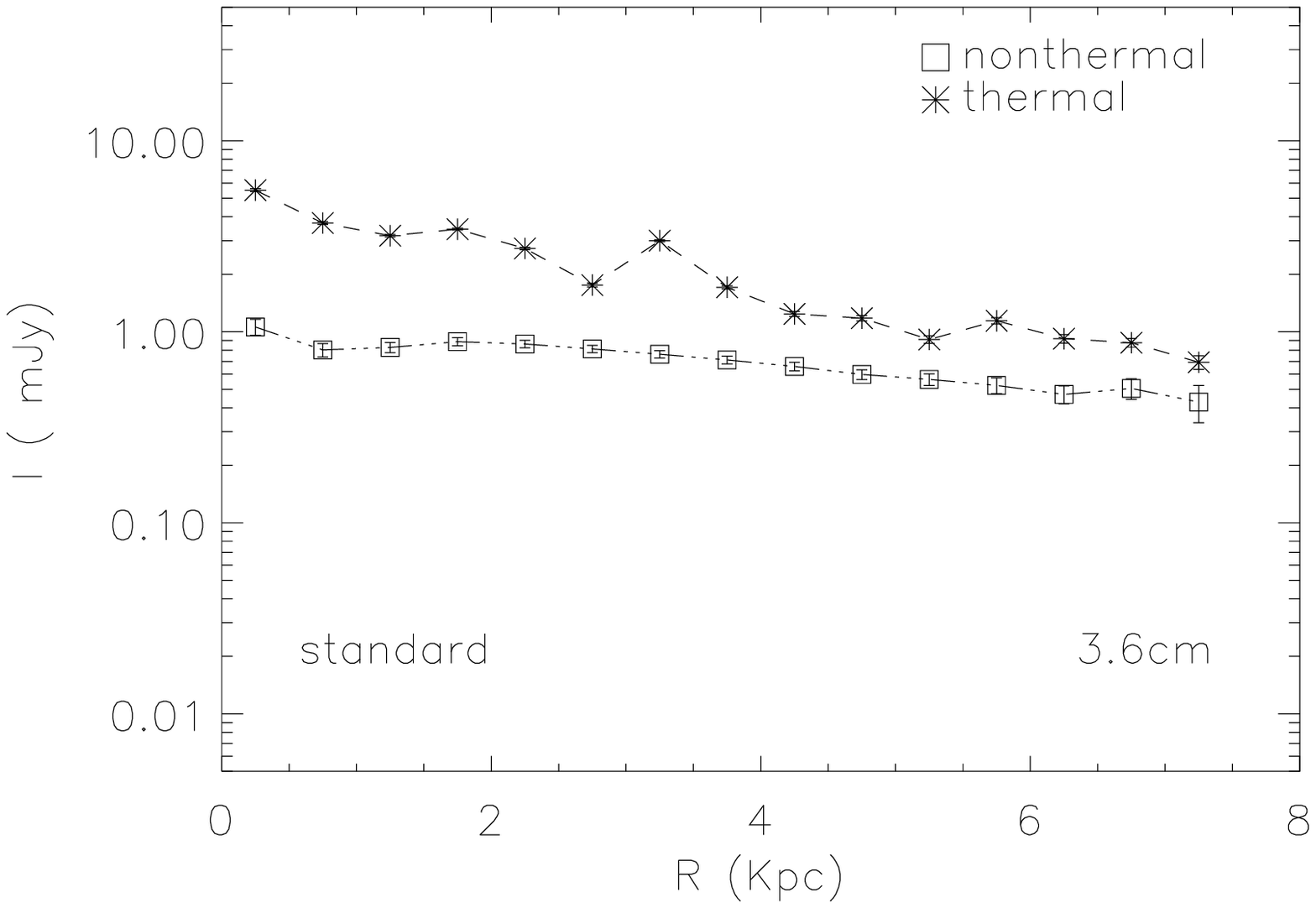}
\includegraphics*{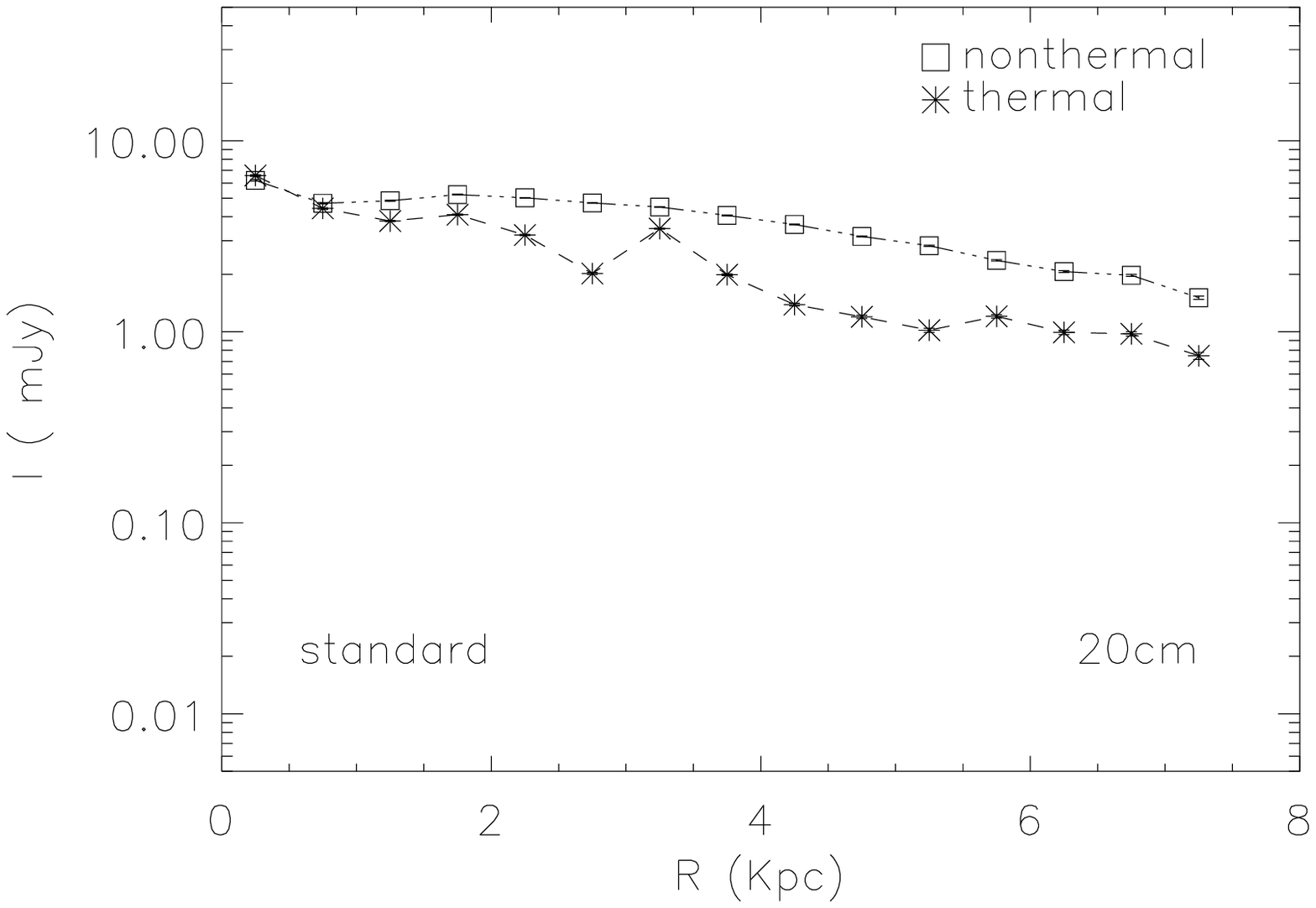}}
\caption[]{Radial profiles of the flux densities of the thermal and nonthermal emission at 3\,cm (left) and 20\,cm (right) obtained from the {\it standard} separation method.  The thermal flux densities show a bump at $3<$\,R\,$<4$\,kpc which is due to the HII complex NGC604.}
\label{fig:surf20cm}
\end{figure*}

Table~4 lists the thermal fractions of the 11 bright HII regions and of M33 (from the integrated flux density maps up to R\,=\,7.5\,kpc) at 3.6\,cm obtained from both methods. The thermal fractions of all the sources are larger when obtained from the standard method than from the new method. Some even exceed 100$\%$, causing very weak or negative nonthermal fluxes. The thermal fraction of M33 from the standard method is 23$\%$\,$\pm$\,14$\%$ higher than that from the new method. In Paper~I, we used the standard method with  $\alpha_n=1.0 \pm 0.1$ to estimate the thermal fraction from the integrated spectrum, based on data at 35.6, 21.1, 17.4, 11.1, 6.3, 6.2, 3.6, and 2.8\,cm, as  0.49\,$\pm$\,0.15 which agrees with the value 0.51\,$\pm$\,0.04
obtained from the new method. This indicates that the assumption of a constant nonthermal spectral index is reasonable to estimate mean values for global studies, when the integrated spectrum is used.

\begin{table*}
\begin{center}
\caption{Thermal fractions at 3.6\,cm obtained from the new and standard methods.}
\begin{tabular}{ l l l l } 
\hline \hline
Object &  $\alpha_{(3.6,20)}$ &  $F_{th}$ (new & $F_{th}$ (standard \\
     &  & method) $\%$ & method) $\%$ \\
%   &  &       $\%$     &      $\%$             \\
\hline
NGC604 & 0.12\,$\pm$\,0.01  & 76.4\,$\pm$\,0.3  &  99.2\,$\pm$\,0.4       \\
NGC595  & 0.07\,$\pm$\,0.03  & 67.6\,$\pm$\,1.3  &  101.6\,$\pm$\,1.4       \\
IC133  & -0.30\,$\pm$\,0.03  & 29.3\,$\pm$\,0.8  &   106.5\,$\pm$\,1.6    \\
B690   & 0.14\,$\pm$\,0.05 &  68.9\,$\pm$\,3.9 &    99.1\,$\pm$\,3.6    \\
B61/62 & 0.04\,$\pm$\,0.08 & 60.2\,$\pm$\,3.7  &    104.3\,$\pm$\,4.2      \\
IC132 & 0.07\,$\pm$\,0.02  & 66.5\,$\pm$\,1.0  &    101.6\,$\pm$\,1.5     \\
IC131  & -0.06\,$\pm$\,0.03& 71.0\,$\pm$\,2.2  &     107.2\,$\pm$\,2.4    \\
NGC588 & 0.00\,$\pm$\,0.05 & 69.8\,$\pm$\,2.6  &   103.1\,$\pm$\,2.9    \\
IC142 & 0.09\,$\pm$\,0.04  & 77.8\,$\pm$\,4.6  &   96.9\,$\pm$\,3.6    \\
B691 & 0.27\,$\pm$\,0.05   &  51.1\,$\pm$\,2.6 &    93.0\,$\pm$\,3.4    \\
NGC592 & 0.13\,$\pm$\,0.04 & 64.5\,$\pm$\,2.6  &   97.9\,$\pm$\,2.8    \\
\hline
M\,33 : & &\\
R$<7.5$\,kpc &  0.72\,$\pm$\,0.04  &51.4\,$\pm$\,4.2    &   63.2\,$\pm$\,5.3      \\
%(mJy) & & & &\\
\hline
\end{tabular}
\end{center}
\end{table*}

\subsection{Radial scale lengths}

Fig.~\ref{fig:surf3cm} shows the exponential distributions of the thermal and nonthermal intensities  at 3.6 and 20\,cm with galactocentric radius. The radial profiles of the nonthermal emission are smoother and flatter than those of the thermal emission at both wavelengths. The fluctuations in the thermal profiles  at $2<$\,R\,$<4$\,kpc cause similar fluctuations observed in the profile of the total emission \citep[presented in ][]{Tabatabaei_2_07}. 

The same profiles obtained from the standard method are shown in Fig.~\ref{fig:surf20cm} for comparison. The nonthermal radial profiles are slightly smoother (especially at 3.6\,cm) than those from the new method, caused by the assumption of a constant nonthermal spectral index. Although the thermal radial profiles exhibit the same variations as those in Fig.~\ref{fig:surf3cm}, the over-estimation of the thermal emission from the standard method is obvious.

The exponential scale length, $l$, is obtained by fitting an exponential function of the form 
$I({\rm R})= I_{0}exp(-{\rm R}/l)$, where $I_0$ is the intensity at R\,=\,0.
Table~5 shows the exponential scale lengths of the thermal emission, $l_{th}$, and nonthermal emission, $l_{n}$, from both methods. 
Generally, the scale lengths of the nonthermal emission are larger than those of the thermal emission (by a factor of $\simeq$\,2). In Sect.~7, we show  that the CR sources follow the distribution of the thermal sources, i.e. star-forming regions. Hence, the radial distribution of the CR sources decreases faster than that of the synchrotron emission. This is a direct observational result indicating diffusion of the cosmic rays from their  places of origin to larger distances. 

%Similar differences in the radial distributions have also been found between the CR sources and $\gamma$-ray emission in the Milky Way. The $\gamma$-ray observations provide information about the spatial CR distribution \citep[e.g. through a model by ][]{Biermann_01}. The EGRET  observations of our Galaxy  showed that the surface density of the CR sources such as supernova remnants  decreases a factor of two faster with Galactocentric radius than the $\gamma$-ray emissivity \citep{Strong_96}. The difference between these radial dependencies has been explained  by cosmic ray propagation effects using self-consistent Galactic wind simulations \citep{Breitschwerdt}. 

%. Subtracting from the total radio 6.2\,cm emission, they also obtained distribution of the nonthermal emission. Then \cite{Berkhuijsen_85} derived a small scale length of $l=1.2 \pm 0.1$\,kpc for the thermal and  $l=2.0 \pm 0.3$\,kpc for the nonthermal emission at $1<R<6$\,kpc. This can be explained by their inability to detect all of the diffuse emission. 
Using the standard method, \cite{Buczilowski_88} determined the scale length of the 6.3\,cm thermal and nonthermal emission as $1.8 \pm 0.2$\,kpc and $4.2 \pm 0.3$\,kpc, respectively.
% for $R<9.6$\,kpc. 
These scale lengths are smaller than those obtained here (even smaller than those obtained from the standard method), although the ratio of the nonthermal to thermal scale lengths is the same ($\simeq$\,2). Due to the low signal-to-noise ratio of the old 6.3\,cm Effelsberg receiver used by \cite{Buczilowski_88} much of the diffuse emission in the outer parts of the galaxy had been missed, leading to steeper radial profiles and smaller scale lengths. For the same reason also the scale lengths derived by \cite{Berkhuijsen_85} are too small. They obtained the distribution of the thermal emission at 6.2\,cm from a catalogue of HII regions in H$\alpha$ \citep{Boulesteix} where the diffuse emission was not completely included.

In case of equipartition between the magnetic field and CRs, the scale length of the CR electrons is given by $l_{cr}= l_{n}(3+\alpha_{n})/2$ and that of the magnetic field by $l_{\rm B}= 2\,l_{cr}$ \citep[e.g. ][]{Klein_82}.
Taking  $\alpha_{n} \simeq 1$, we obtain $l_{cr}\simeq 12$\,kpc and  $l_{\rm B}\simeq 24$ \,kpc. 

For NGC6946, \cite{Walsh} found a nonthermal scale length of $l_{n} \sim 4$\,kpc which gives a smaller CR scale length of $l_{cr}\simeq 8$\,kpc. Although NGC6946 is a Scd-type galaxy like M33, it is a starburst system and it hosts stronger star formation in its central region than M33. This may cause a steeper radial profile of the total intensity and consequently a smaller nonthermal (and cosmic ray) scale length than in M33. 

%It is possible that their large beam of 7.7$\arcmin$ (1.85\,kpc) influenced the scale length estimation besides the fact that our study is limited to smaller radii ($R<7.5$\,kpc).

\begin{table}
\begin{center}
\caption{Exponential scale lengths of the thermal and nonthermal components of the radio contiuum emission from M33.}
\begin{tabular}{ l l l } 
\hline
$\lambda$ & $l_{th}$\,(kpc) & $l_{n}$\,(kpc)\\
\hline 
new method & & \\
\hline
20\,cm & 2.4\,$\pm$\,0.2  & 5.8\,$\pm$\,0.5\\
3.6\,cm& 2.6\,$\pm$\,0.2   & 6.1\,$\pm$\,0.7 \\
\hline
standard method & & \\
\hline
20\,cm & 3.5\,$\pm$\,0.5  & 6.2\,$\pm$\,0.7\\
3.6\,cm& 3.7\,$\pm$\,0.5   & 8.9\,$\pm$\,0.9 \\
\hline
\end{tabular}
\end{center}%\label{tab:scalelength}
\end{table}

\subsection{North--south asymmetry}

%\cite{Tabatabaei_2_07} find a north-south (N-S) asymmetry in polarization that is wavelength-dependent, indicating an asymmetry in Faraday depolarization. As the thermal electrons are responsible for Faraday depolarization, comparing the distribution of the thermal emission in the northern and southern halves of the galaxy may give an insight to this asymmetry. 
From Figs.~\ref{fig:ther3.6cm} and \ref{fig:ther20cm} it seems that the thermal emission is stronger in the southern than in the northern half of M33. To investigate this north-south (N-S) asymmetry, we obtain the integrated flux density of the thermal, nonthermal, and total emission in each half separately. Table~6 shows the results from both the new and the standard method.  The thermal emission from the new method is slightly stronger in the southern than in the northern half of the galaxy at both wavelengths. This may be the reason for the higher Faraday depolarization found in the southern half than in the northern half \citep{Tabatabaei_2_07}.  In Paper III, we will discuss whether the asymmetry in the Faraday depolarization can be caused by this N-S asymmetry in the thermal emission distribution. Note that, from the standard method, the thermal emission in the southern half is weaker. 

The nonthermal emission at 20\,cm from the new method in the southern half is also stronger than in the northern half (in contrast to the 3.6\,cm nonthermal emission). This is expected as there are more supernova remnants in the southern half, $\simeq\,56\%$ of the total number, especially in the main southern arm IS (see Fig.~\ref{fig:20cmsnr}), which produce stronger nonthermal emission at longer wavelengths. The existence of the strong optically thick HII regions (e.g. IC133) in the northern half of the galaxy also causes the different N-S ratios of the nonthermal emission at 3.6\,cm.

\begin{table*}
\begin{center}
\caption{ North-South ratios of the integrated flux densities of the thermal, nonthermal, and total radio emission.}
\begin{tabular}{ l l l l} 
\hline
$\lambda$\,(cm)& S$_{th}^{N}$/S$_{th}^{S}$ & S$_{n}^{N}$/S$_{n}^{S}$ & S$_{T}^{N}$/S$_{T}^{S}$\\
\hline 
\hline
new method & & & \\
\hline
3.6 &  0.89\,$\pm$\,0.05 & 1.15\,$\pm$\,0.08 & 1.10\,$\pm$\,0.08  \\
20 & 0.86\,$\pm$\,0.05 & 0.75\,$\pm$\,0.07 & 0.82\,$\pm$\,0.07 \\
\hline
standard method & & & \\
\hline
3.6 &  1.20\,$\pm$\,0.07 & 0.67\,$\pm$\,0.07 & 1.10\,$\pm$\,0.08 \\
20 & 1.14\,$\pm$\,0.06 & 0.77\,$\pm$\,0.08 & 0.82\,$\pm$\,0.07 \\
%\hline
%\hline
%$\mu$m & (Jy) &(Jy)&&&&&&&\\
%\hline
%24\, & 18\,$\pm$\,2& 17\,$\pm$\,1 &1.07\,$\pm$\,0.12 &&&&&\\
%70 & 265\,$\pm$\,9 & 291\,$\pm$\,11&0.91\,$\pm$\,0.05 &&&&&\\
%160 &743\,$\pm$\,28 & 862\,$\pm$\,34&0.86\,$\pm$\,0.05 &&&&&\\

\hline
\end{tabular}
\end{center}
\end{table*}

\subsection{Uncertainties}

How do the assumptions of the new method of separating thermal and nonthermal emission influence the results? We take the thermal fraction (e.g. at 3.6\,cm) as the final result and define $U$ as the uncertainty in $F_{th}$ when one of the assumptions changes,
\begin{equation}
U \equiv \frac{\vert F_{th} - F'_{th}\vert}{F_{th}}.
\end{equation}
$F_{th}$ and $F'_{th}$ are the thermal fractions before and after changing an assumption, respectively.

One of the assumptions is the choice of effective extinction factor $f_d$\,=\,0.33 (non-uniform ionization).  For a homogeneous distribution of dust and ionized gas, $f_d$=\,0.5, the average uncertainty in the thermal fraction at 3.6\,cm is only $U=2\%$ with a standard deviation of 1$\%$.

\begin{figure}
\resizebox{\hsize}{!}{\includegraphics*{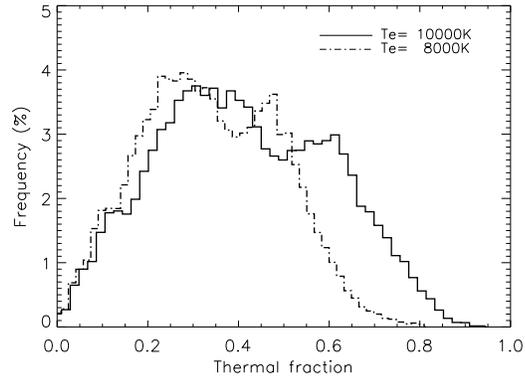}}
\caption[]{Distribution (histogram) of the thermal fraction at 3.6\,cm in M33. The number of bins used for this histogram is 50.  }
\label{fig:Fth}
\end{figure}

Changing $T_e$ from the adopted 10000\,K to 8000\,K, the most probable uncertainty across the galaxy is  $U=17\%$  with  a standard deviation of 3$\%$. Fig.\ref{fig:Fth} shows that the thermal fraction decreases with decreasing electron temperature. Larger differences in the thermal fraction are found for regions with higher thermal emission like HII regions, and the influence of changing $T_e$ is relatively small in diffuse regions with small $F_{th}$.  Thus, we find that the uncertainty resulting from the new method is mainly determined by uncertainty of the electron temperature.

Another question is which method,  the new method with its electron temperature assumption or the standard method with its constant nonthermal spectral index assumption, faces larger uncertainty in the thermal fraction.
For this comparison, we first obtain the nonthermal spectral indices for different electron temperatures. For  $T_e$\,=10000\,K the most probable nonthermal spectral index is 0.9, while it is 0.8 for $T_e$\,=8000\,K. Then, from the standard method, we calculate the thermal fractions assuming $\alpha_{n}$\,=\,0.9 and 0.8, respectively, which leads to an uncertainty of $U=40\%$, much larger than that determined from our method (17$\%$). We conclude that the thermal fraction is more sensitive to variations of the nonthermal spectral index in the standard method than to variations of the electron temperature in the new method. 

Table~7 shows the thermal fractions obtained from the thermal and total flux densities of M33 (integrated for R\,$<$\,7.5\,kpc) for different combinations of the assumptions of $T_e$ and $f_d$.

%Using a cross-correlation study of the K- and Ka-band WMAP data, \cite{Davis_06} found an electron temperature of 4000-5000\,K for diffuse ionized gas at the intermediate Galactic latitudes.  Assuming $T_e$\,=4000\,K for M33 leads to low thermal fractions: at 3.6\,cm the thermal fraction would be only $\sim$\,38$\%$ for NGC604 (with $\alpha$\,=\,0.12) and $<\,10\%$ for diffuse emission. Therefore, the electron temperature attributed to the Galactic diffuse emission ($T_e$\,=4000\,K) seems to be not applicable for M33, particularly, as the free-free emission is dominated by the disk of the galaxy. 
%\cite{Paladini} found a thermal fraction of 40\,-\,82$\%$ for the Galactic coordinates of $20^{\circ}\, <\,l\,<\,30^{\circ}$ and $-1.^{\circ}5\, <\,b\,<\,+1.^{\circ}5$ at 5\,GHz. This is equivalent to a mean nonthermal spectral index of $\geq$\,0.9 for M33 (with the mean total spectral index of $\sim$\,0.7) leading to an electron temperature of $\geq$\,10000\,K for M33. Thus our assumption of  $T_e$\,=10000\,K gives the lower limit of thermal fraction for the warm ionized gas given by \cite{Paladini}.

%\begin{figure*}
%\begin{center}
%\resizebox{15cm}{!}{\includegraphics*{spectoo.ps}
%\includegraphics*{spectss.ps}}
%\caption[]{ The wavelet spectra of the 20 and 3.6\,cm images at $90\arcsec$ resolution before (left) and after (right) source subtraction. The data points correspond to the scales 90, 143, 183, 234, 299, 383, 490, 626, 800, $938\arcsec$. The spectra are shown in arbitrary units.  }
%\end{center}
%\end{figure*}

\begin{table}
\begin{center}
\caption{Average thermal fraction of M33 at 3.6\,cm for different assumptions. The uncertainty $U$ of each assumption is calculated with respect to the case $T_e$\,=\,10000\,K and $f_d$\,=\,0.33. }
\begin{tabular}{ l l l} 
\hline
\hline
 \,\,\,\,\,$T_e$\,\,\,\,\,\,\,\,\,\& \,\,\,\,\,\,$f_d$ &\,\,\,\,\,\,\,\,\,\,\,\,\,\,\,\,\,\,\,\,\,$F_{th}$  &\,\,\,\,\,\,\,\,\,\,\,\,\,\,\,\,\,\,\,\, $U$    \\
  \,\,\,(K)\,\,\,\,\,\,\,\,\,\,\,\,\,\,\,\,\,\,\,(...)   & \,\,\,\,\,\,\,\,\,\,\,\,\,\,\, \,\,\,(\%)   &\,\,\,\,\,\,\,\,\,\,\,\,\,\,\,\,\,\,\,\,(\%)   \\
\hline
10000\,\,\,\,\,\&\,\,\,\, 0.33 & \,\,\,\,\,\,\,\,\,\,\,\,\,\,\,\,\,\,\,\, 51  &  \,\,\,\,\,\,\,\,\,\,\,\,\,\,\,\,\,\,\,\,(...) \\
10000\,\,\,\,\,\&\,\,\,\, 0.50  & \,\,\,\,\,\,\,\,\,\,\,\,\,\,\,\,\,\,\,\, 54  & \,\,\,\,\,\,\,\,\,\,\,\,\,\,\,\,\,\,\,\,\,\, 5         \\
\,\,\,8000\,\,\,\,\,\&\,\,\,\, 0.33  & \,\,\,\,\,\,\,\,\,\,\,\,\,\,\,\,\,\,\,\, 41  & \,\,\,\,\,\,\,\,\,\,\,\,\,\,\,\,\,\,\, 20 \\
\,\,\,8000\,\,\,\,\,\&\,\,\,\, 0.50  & \,\,\,\,\,\,\,\,\,\,\,\,\,\,\,\,\,\,\,\,\,\,43   & \,\,\,\,\,\,\,\,\,\,\,\,\,\,\,\,\,\,\, 16\\ 
\hline
\end{tabular}
\end{center}%\label{tab:scalelength}
\end{table}

\section{Summary and conclusions}

We have developed a new method to separate the thermal and nonthermal radio emission from a galaxy. We used the highly resolved and sensitive Spitzer 70 and 160\,$\mu$m data of M33 to correct the H$\alpha$ map of \cite{Hoopes_et_al_97H} for extinction. From this map, we calculated the thermal (free-free) emission at 3.6 and 20\,cm and obtained maps of the nonthermal emission as well as a map of the nonthermal spectral index in M33. The distribution of the nonthermal spectral index greatly  helps to understand the origin and propagation of cosmic ray electrons in a galaxy. In brief, the results and conclusions are as follows:
\begin{itemize}
\item[$\bullet$] The distribution of the dust extinction is similar to that of the 160\,$\mu$m emission. The mean extinction in rings in the galactic plane exhibits a shallow radial gradiant.
    
\item[$\bullet$] With a nonthermal fraction of about 30\%--60\% at 3.6\,cm, the spiral arms and star-forming regions have a considerable contribution to the nonthermal emission. This contribution is negligible in the nonthermal maps obtained from the standard separation method. The radial profiles of the surface brightnesses and the wavelet spectra show that the distribution of the nonthermal emission from the standard method is smoother than that derived from the new method. This is caused by the assumption of a constant nonthermal spectral index in the standard method.

\item[$\bullet$] The nonthermal emission from the new method is still more smoothly distributed than the thermal emission. The exponential scale lengths of the nonthermal emission are more than twice as large as those of the thermal emission.

\item[$\bullet$] The standard method over-estimates the thermal fraction, especially at the position of giant HII regions. For galactocentric radius R$<$\,7.5\,kpc, the thermal fractions at 3.6\,cm are 51\,$\pm$\,4$\%$ and 63\,$\pm$5\,$\%$ from the new and standard methods, respectively.

\item[$\bullet$] For the first time, we derived a map of the nonthermal spectral index. In the star-forming regions, the nonthermal spectrum is relatively flat with an average value of $\alpha_n$ of 0.6\,$\pm$\,0.1, the typical spectral index of supernova remnants, but $\alpha_n$ increases to $1.2 \pm 0.2$ in the interarm regions and outer parts of the galaxy. This shows that the relativistic electrons lose energy when diffusing from their origin in star-forming regions towards interarm regions and the outer parts of the galaxy. The mean spectral index of the nonthermal emission becomes equal to that of the total emission at R\,$\simeq$\,4.5\,kpc. This indicates that the total radio emission is mostly nonthermal at  R\,$>$\,4.5\,kpc in M33, where the spectral index is dominated by synchrotron and inverse Compton loss processes.  

\item[$\bullet$] The north-south asymmetry in the distribution of the thermal emission obtained from the new method is opposite to that from the standard method. The thermal fractions are 47\,$\pm\,5\%$ and 53\,$\pm\,5\%$ in the northern and southern half, respectively. 

\item[$\bullet$] Generally, the {\it integrated} results from the two methods match with each other within the errors, indicating that the assumption of a constant nonthermal spectral index is a proper approximation for `global' studies.
 
%This is seen particularly in the mean nonthermal spectral index in rings, where there is no general increasing or decreasing trend with galactocentric radius (Sect. 8). That is also indicated by a compromise between the total thermal fraction from the new method and that obtained from the integrated spectrum (Sect. 9.1).  
\end{itemize}

%Finally, this developed method is also applicable for other face on galaxies for which the data of dust emission exist.

\begin{acknowledgements}
We are grateful to P. L. Biermann for valuable and stimulating comments. 
We thank the staff of the 100--m Effelsberg telescope and the VLA  for their assistance with radio observations.  
F. Tabatabaei was supported for this research through a stipend from the International Max Planck Research school (IMPRS) for Radio and Infrared Astronomy at the Universities of Bonn and Cologne. 
We acknowledge the M33 Spitzer collaboration (PI: R. Gehrz) for the MIPS observations.
\end{acknowledgements}

\bibliography{s.bib}        % rezabib.bib is name of the database
\end{document}